\documentclass[journal]{IEEEtran}

\usepackage{hyperref}

\usepackage{array}
\newcolumntype{P}[1]{>{\centering\arraybackslash}p{#1}}

\ifCLASSINFOpdf
\else
   \usepackage[dvips]{graphicx}
\fi
\usepackage{url}

\usepackage{graphicx}
\usepackage{color}
\usepackage{xcolor}
\usepackage{amsmath}
\usepackage[flushleft]{threeparttable}
\usepackage{subcaption}
\usepackage{cite}

\begin{document}

\title{Gradient-based Adversarial Deep Modulation Classification with Data-driven Subsampling}

\author{Jinho Yi, \IEEEmembership{Student Member, IEEE}, and Aly El Gamal, \IEEEmembership{Senior Member, IEEE}
\thanks{J. Yi, and A. El Gamal are with the Department of Electrical and Computer Engineering, Purdue University, West Lafayette, IN, USA. Email: \{yi62, elgamala\}@purdue.edu.}}

\maketitle

\begin{abstract}
Automatic modulation classification can be a core component for intelligent spectrally efficient wireless communication networks, and deep learning techniques have recently been shown to deliver superior performance to conventional model-based strategies, particularly when distinguishing between a large number of modulation types. However, such deep learning techniques have also been recently shown to be vulnerable to gradient-based adversarial attacks that rely on subtle input perturbations, which would be particularly feasible in a wireless setting via jamming. One such potent attack is the one known as the Carlini-Wagner attack, which we consider in this work. We further consider a data-driven subsampling setting, where several recently introduced deep-learning-based algorithms are employed to select a subset of samples that lead to reducing the final classifier's training time with minimal loss in accuracy. In this setting, the attacker has to make an assumption about the employed subsampling strategy, in order to calculate the loss gradient. Based on state of the art techniques available to both the attacker and defender, we evaluate best strategies under various assumptions on the knowledge of the other party's strategy. Interestingly, in presence of knowledgeable attackers, we identify computational cost reduction opportunities for the defender with no or minimal loss in performance. 

\end{abstract}

    
\IEEEpeerreviewmaketitle

\section{Introduction}
Automatic modulation classification (AMC) lets the receiving device recognize the modulation type of the received signal with no manual engineering or prior agreement with the transmitter side. This could be one of the core functions of next generation intelligent spectrally efficient networks, which allows communication devices to freely adapt to various wireless systems, not compromising on the spectrum efficiency. To facilitate AMC, gradient-based Machine Learning (ML) techniques with successful track record in computer vision \cite{yolo} and natural language processing \cite{bert} have been recently investigated, showing a promising performance \cite{O_Shea_2016, 8267032}. The ML approach showed superior performance to conventional modulation classification algorithms, that are based on statistical heuristics, in a scalable fashion with both numbers of modulation types and samples. In order to address the feasibility of this promising technology, we consider in this work two major aspects: Robustness against a malicious hamper and computational efficiency.  

\subsection{Gradient-based Adversarial Machine Learning}
ML models, that rely on stochastic gradient descent optimization of deep neural networks, are known to lack robustness against malicious adversarial examples \cite{intriguingProperties, goodfellow2014explaining}, which introduce small perturbations, specially crafted to cause ML models to malfunction when added to legitimate inputs. This perturbation is obtained by solving an optimizing problem that minimizes the perturbation norm while maximizing the classification error. In a wireless setting, a small perturbation norm would correspond to small transmit power needed by a jammer to introduce slight additional noise in the signal, that is difficult to detect at the receiver and leads to significant ML performance degradation. Recent studies confirmed the vulnerability of ML models for modulation classification to such adversarial examples. Work by Usama et al. \cite{8881843, usama2019blackbox} have demonstrated that the famous CW attack \cite{Carlini_2017} is applicable in this setting. Furthermore, Kim et al. \cite{Kim_2020}, Hameed et al. \cite{8969541}, and Bair et al \cite{10.1145/3324921.3328785} have presented effective algorithms to generate adversarial examples, which take in account the noise encountered during actual over-the-air transmission of the perturbed signals. 

\subsection{Data-driven Subsampling}
ML models also challenge the computational power of a wireless device, particularly during the training phase. The long training time of a deep neural network causes a severe bottleneck for its application in wireless communications, where frequent re-training would be needed to adapt to the varying environment in real-time. Ramjee et al. \cite{ramjee2020ensemble}, therefore, introduced a data-driven subsampling strategy, through simulations employing deep learning models, called \emph{Subsampler Nets}, to sample down the input to the ML model and effectively reduce the size of the deep neural network architecture and its training time. Their algorithm exploits the transferability property in deep learning, which capitalizes on having common relevant features among various architectures that are fit for the same task. Using multiple pre-trained modulation classifiers, the algorithm ranks individual input samples on how much each contributes to the classification outcome.

\subsection{Considered Problem}
Motivated to address the aforementioned two aspects, we study the robustness against adversarial attacks of an effective ML-based AMC model while employing different data-driven subsampling strategies. As we show in the sequel, knowledge of the employed Subsampler Net can dramatically affect the attacker's choice of best strategy and its effectiveness. Similarly, knowledge of the assumed Subsampler Net by the attacker significantly impacts the defender's strategy. We finally identify a computationally efficient training strategy with minimal cost in performance for a system whose design is perfectly exposed to the attacker.

\section{Problem Setup}
We consider a deep neural network \textit{victim classifier} for modulation classification and an \textit{attacker} that attempts to perturb the classifier's input by jamming the wireless transmit signal.
We have investigated the performance of \textit{victim} models, trained with different data-driven subsampling strategies, with and without \textit{attacker's} perturbations. The following subsections describe how we set these experiments in detail. 

\subsection{Dataset for Modulation Scheme}

We have used the RML2016.10b dataset generated with GNU radio \cite{grcon}, which is highly cited and open source, enabling reproducibility of our results\footnote{Source code for this work has been accepted for publication at\\ https://codeocean.com/capsule/8397297/tree/v1. We used Keras with Tensorflow 2.0 as a backend on a Tesla V100 GPU.}. The dataset consists of received signal complex samples corresponding to 10 different modulation types with uniformly distributed SNR from -20 dB to 18 dB in steps of 2 dB. Each example has 128 samples for in-phase and quadrature (I/Q) components of the signal, represented as a 2 x 128 vector. The 1,200,000 vectors were distributed equally into training and testing examples. 

\subsection{Subsampling Schemes and Modulation Classifier Model}

As mentioned in Section I, we have used Subsampler Nets for sampling down example vectors, targeting the use cases that prioritize computational efficiency such as that of low-power devices. We employed four different Subsampler Net schemes, namely the CNN, CLDNN, ResNet, and the Holistic subsampler, whose implementation details are in \cite{ramjee2020ensemble}. The Holistic subsampler sorts out the best samples from the sample sets selected by the three Subsampler Nets based on CNN, CLDNN and ResNet models. Only the training data set was used for selecting the sample indices. Furthermore, we used uniform subsampling by sampling at regular intervals, which showed a promising result in \cite{ramjee2019fast}. 

For the classifier, we used a ResNet architecture identical to the one utilized in \cite{ramjee2019fast}, which is initially inspired by \cite{8267032}. It accepts an input of size (1, 2, N) representing (channel, I/Q, number of samples) and utilizes three residual stacks followed by three fully connected layers. A convolutional layer, two residual units, and a max-pooling layer compose each residual stack; each residual unit consists of two convolutional layers with a filter size of 1x5 and a shortcut from the input to the output of the unit. The model is trained using the Adam Optimizer with categorical cross-entropy loss function, a batch size of 1024, and a learning rate of 0.001.

The sample dimension of the training data set was reduced down to 64, 32, and 16 from the original 128, applying each of the five subsampling scheme, and the outputs were used to train 15 distinct \emph{victim classifier} models.

\subsection{Gradient-based Adversarial Attack}
We evaluated the robustness of our classifier against the popular CW $L_2$ Attack \cite{Carlini_2017}, which finds $x$, a perturbed version of the original example $x_0$, through the following optimization:

\begin{equation}\label{cwl2} 
{\underset{x}{\text{minimize}}} \bigg|\bigg|x - x_0\bigg|\bigg|_2 ^2 + c \cdot  f_t(x), \\
\end{equation}

where $f_t$ is defined by:

\begin{equation}\label{eq:cw}
f_t(x') = \text{max}(\text{max}\{Z(x')_i:i \neq t\} - Z(x')_t, 0)
\end{equation}

where t is a target label, and $Z(.)$ is a softmax function. We considered an untargeted attack, where the choice of $t$ minimizing \eqref{eq:cw}, among all labels except the true label, is selected.
We limited the perturbation norm so that its power is the same as the noise power in the final signal.

\subsection{Threat Models}
We considered two attacking scenarios: {\bf1)} a \textit{white-box} adversarial attack where we assumed that the adversary is perfectly knowledgeable about the \textit{victim}'s choice of sub-sampling scheme; {\bf2)} a \textit{black-box} attack where the adversary conjectures what sub-sampling scheme is being used. In both scenarios, the \textit{attacker} knows the victim classifier's architecture and hyperparameters as well as the training dataset.  

To simulate attacking scenarios, we trained surrogate classifiers with the same ResNet model as the victim's to generate adversarial examples, separately from the model used for performance evaluation. We crafted adversarial examples using the testing dataset by employing these surrogate classifiers. 
 For the black-box attack, we averaged each classifier's performance over all the adversarial examples that were crafted to target all possible classifiers of matching subsampling rates.


\section{Experimental Results}\label{sec:results}
First, we review the overall impact of gradient-based adversarial attacks on the classification accuracy while employing different subsampling schemes, then we analyze the best strategies from both the \textit{victim's} and \textit{attacker’s} perspectives. 

\subsection{Performance Impact}
\begin{figure}[ht]
\centerline{\includegraphics[width=8cm]{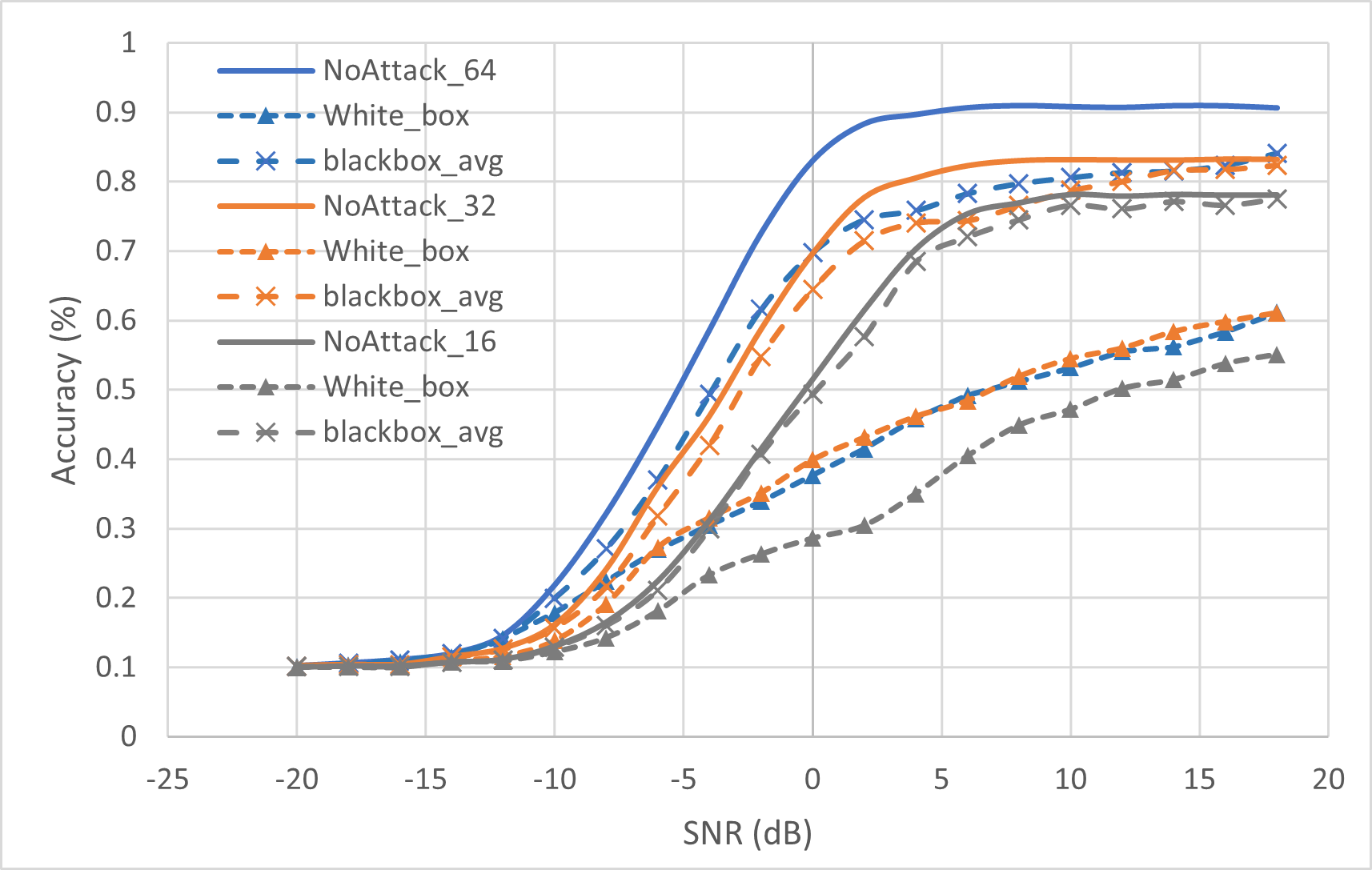}}
\caption{Accuracy vs SNR for ResNet classifier with uniformly selected samples under different attack scenarios with subsampling rates of $\frac{1}{2}$ ($64$ sample input), $\frac{1}{4}$ and $\frac{1}{8}$.}
\label{fig:unifrom}
\end{figure}

Fig. \ref{fig:unifrom} presents a detailed comparison of the accuracy of ResNet classifier models, trained using the inputs sampled with the uniform subsampler, with respect to the proposed threat scenarios. The results for all models with input sizes of $64$ (subsampling rate of $\frac{1}{2}$), $32$, and $16$ are presented in the figure, manifesting their performance drop for both \textit{black-box} and \textit{white-box} attacks. The black-box attack data represents the average accuracy that each model showed across every adversarial example that was targeting the other considered subsampling strategies. This applies henceforward.

\begin{table*}[ht]
\centering\footnotesize
\caption{Classifier accuracy under different attack scenarios}
\label{table:avgAcc}\medskip
\begin{threeparttable}
\begin{tabular}{cccc|ccc|ccc}
\hline
\multicolumn{1}{c|}{Sample \#}       & \multicolumn{3}{c|}{64}       & \multicolumn{3}{c|}{32}  & \multicolumn{3}{c}{16} \\
\multicolumn{1}{c|}{Subsampler Net} & No Attack & Black-box & White-box & No Attack & Black-box & White-box & No Attack & Black-box & White-box \\ \hline
\multicolumn{1}{c|}{CLDNN}          & 56.2\%      & 45.4\%      & 32.8\%      & 50.7\%      & 44.1\%      & 28.7\% & 43.0\%       & 39.8\%      & 26.5\%\\
\multicolumn{1}{c|}{CNN}            & 56.3\%      & 45.7\%      & 29.9\%      & 50.8\%      & 43.1\%      & 29.9\% & 44.5\%       & 39.6\%      & 24.6\%\\
\multicolumn{1}{c|}{ResNet}         & 56.4\%      & 46.6\%      & 33.0\%      & 52.0\%      & 46.1\%      & 29.7\% & 44.9\%       & 41.6\%      & 27.4\%\\
\multicolumn{1}{c|}{Holistic}       & 56.9\%      & 44.4\%      & 33.6\%      & 51.5\%      & 43.6\%      & 28.4\% & 44.9\%       & 39.5\%      & 25.7\%\\
\multicolumn{1}{c|}{Uniform}        & 59.3\%      & 52.1\%      & 34.9\%      & 52.3\%      & 49.3\%      & 35.0\% & 45.2\%       & 44.0\%      & 29.2\%\\ \hline
\end{tabular}
  \begin{tablenotes}[para,flushleft]
  Accuracy of classifiers with different subsampler Nets for no attack, black-box attack, and white-box attack scenarios.
  \end{tablenotes}
\end{threeparttable}
\end{table*}
\begin{table*}
\centering\footnotesize
\caption{\% Decrease in classifier accuracy under different attack scenarios}
\label{table:percentDec}\medskip
\begin{threeparttable}
\begin{tabular}{P{2cm}|P{2cm}P{2cm}|P{2cm}P{2cm}|P{2cm}P{2cm}} 
\multicolumn{1}{c}{} &\multicolumn{6}{c}{\% Decrease in Accuracy} \\
\hline
\multicolumn{1}{c|}{Sample \#} & \multicolumn{2}{c|}{64} & \multicolumn{2}{c|}{32} & \multicolumn{2}{c}{16}\\
\multicolumn{1}{c|}{Subsampler Net} & White-box & Black-box & White-box & Black-box & White-box & Black-box\\ \hline
\multicolumn{1}{c|}{CLDNN} & 41.5\% & 19.1\% & 43.3\% & 13.0\% & 38.4\% & 7.4\% \\
\multicolumn{1}{c|}{CNN} & 46.9\% & 18.9\% & 41.3\% & 15.2\% & 44.7\% & 11.2\%\\
\multicolumn{1}{c|}{ResNet} & 41.4\% & 17.4\% & 42.9\% & 11.3\% & 39.1\% & 7.5\%\\
\multicolumn{1}{c|}{Holistic} & 41.0\% & 21.9\% & 44.9\% & 15.4\% & 42.7\% & 12.0\%\\
\multicolumn{1}{c|}{Uniform} & 41.1\% & 12.2\% & 33.2\% & 5.7\% & 35.4\% & 2.7\%\\ \hline
\end{tabular}
  \begin{tablenotes}[para,flushleft]
  Percent decrease ($\Delta acc / acc_{orig}$) in the accuracy of classifiers with different subsampler Nets for black-box and white-box attack scenarios.
  \end{tablenotes}
\end{threeparttable}
\end{table*}
Note that there is a substantial difference in  the relative accuracy drop due to \textit{white-box} and \textit{black-box} attacks, considering that the attacker is fully knowledgeable about the \textit{victim classifier}'s architecture in both cases and the difference in attacker's knowledge between the two attacks is only about the subsampling strategy. 
Table \ref{table:avgAcc} indicates that such performance gap between \textit{black-box} and \textit{white-box} settings is consistently observed even when employing different subsampling schemes than uniform subsampling. It is worth noting how in the presence of attacks, the performance using subsampling rates of $\frac{1}{2}$ and $\frac{1}{4}$ is similar across the whole considered SNR range. This indicates an opportunity for saving computational and hardware costs of systems deployed in significantly adversarial environments, through aggressive subsampling rates. Further, we note how the effectiveness of the black box attack decreases with more aggressive subsampling rates; that it has no noticeable effect for a subsampling rate of $\frac{1}{8}$. We can find this more evidently in Table II, which shows that the percent decrease in accuracy due to the black-box attack is only $2.7\%$ when using 16 input samples. This also signifies the potential of subsampling as the performance losses due to computational efficiency gains and adversarial attacks do not add up.
\begin{figure}[ht]
 \centering
 \begin{subfigure}[a]{8cm}
     \centering
     \includegraphics[width=8cm]{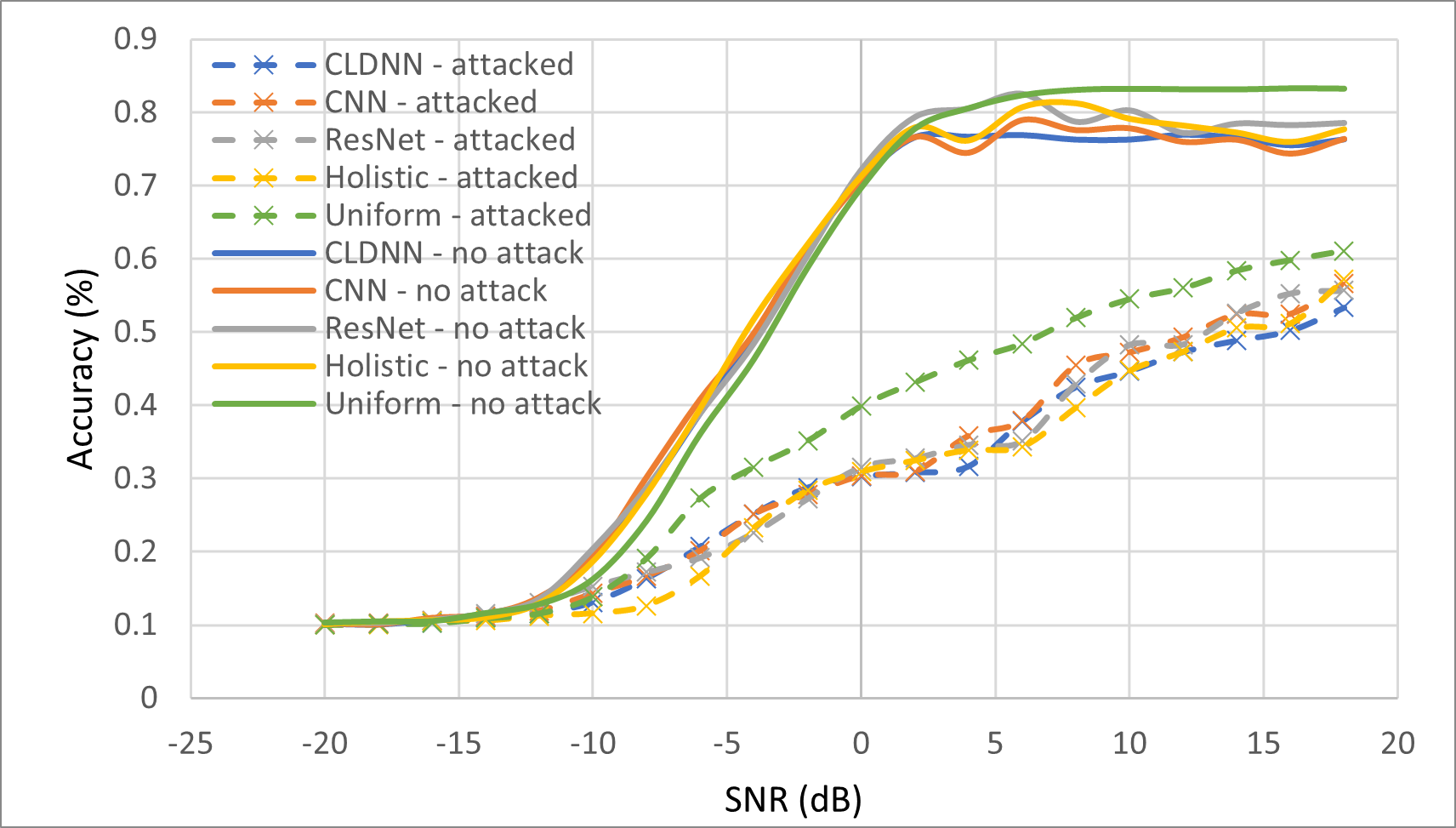}
     \caption{Accuracy vs SNR for 32 sample classifiers before and after the white-box adversarial attack.}
     \label{fig:white32abs}
 \end{subfigure}
 \hfill
 \begin{subfigure}[b]{8cm}
     \centering
     \includegraphics[width=8cm]{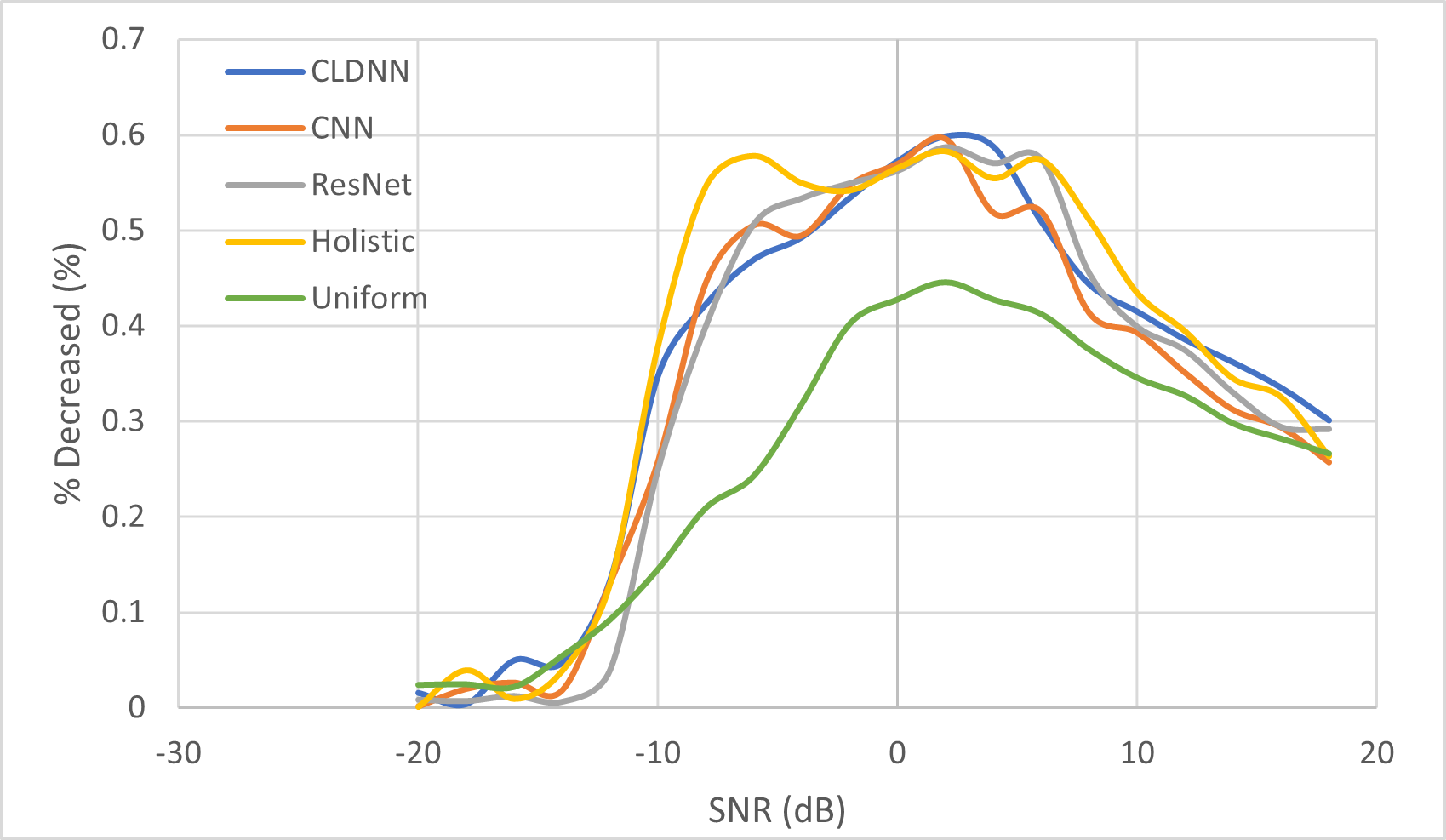}
     \caption{Percent decrease in accuracy due to the white-box adversarial attack vs SNR for 32 sample classifiers.}
     \label{fig:white32pec}
 \end{subfigure}
 \hfill
    \caption{Classification accuracy of different classifiers of 32 input samples in presence of a \textit{white-box} adversarial attack.}
    \label{fig:whitebox}
\end{figure}
\begin{figure}
 \centering
 \begin{subfigure}[a]{8cm}
     \centering
     \includegraphics[width=8cm]{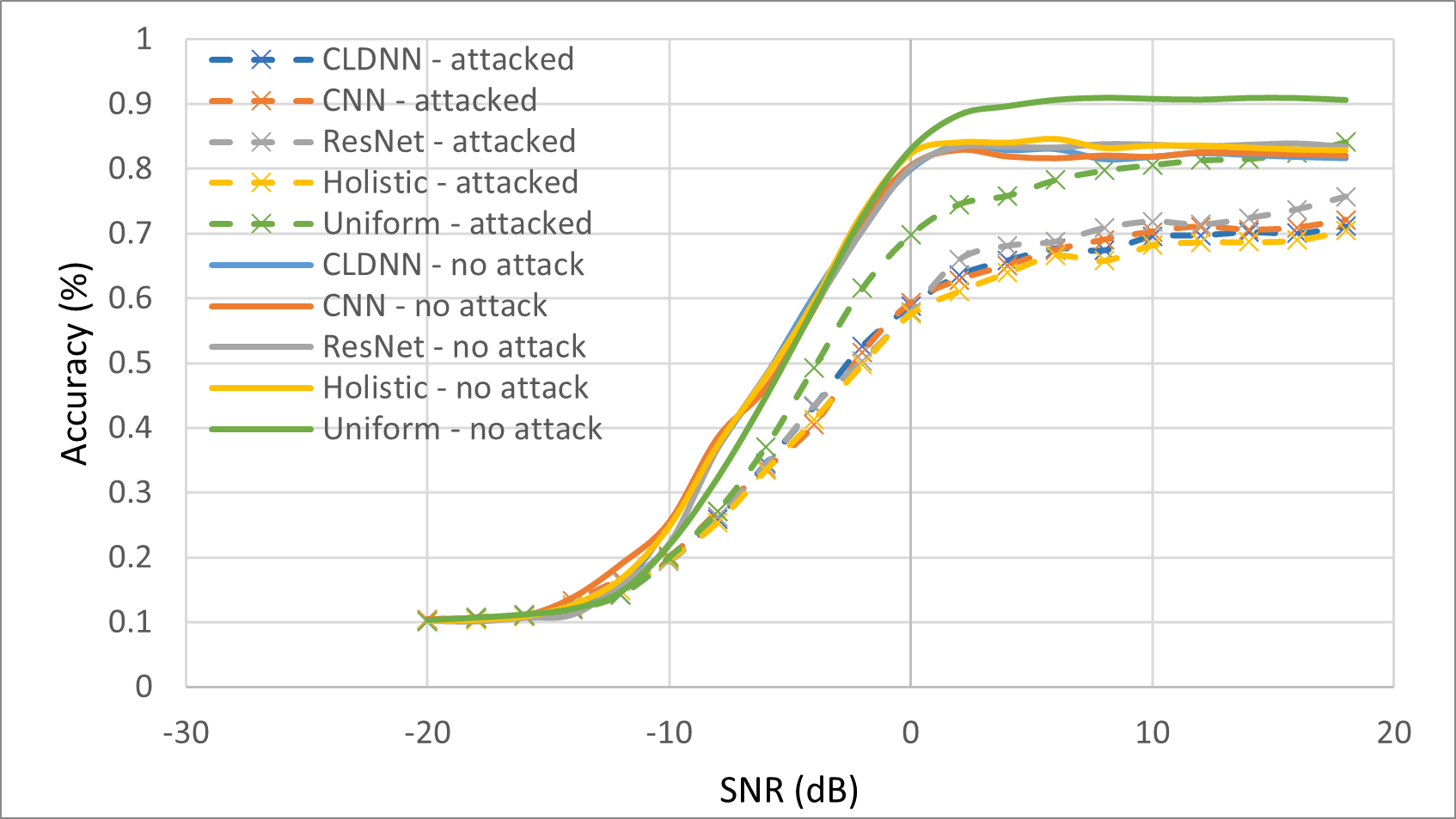}
     \caption{Accuracy vs SNR for 64 sample classifiers before and after the black-box adversarial attack.}
     \label{fig:black64abs}
 \end{subfigure}
 \hfill
 \begin{subfigure}[b]{8cm}
     \centering
     \includegraphics[width=8cm]{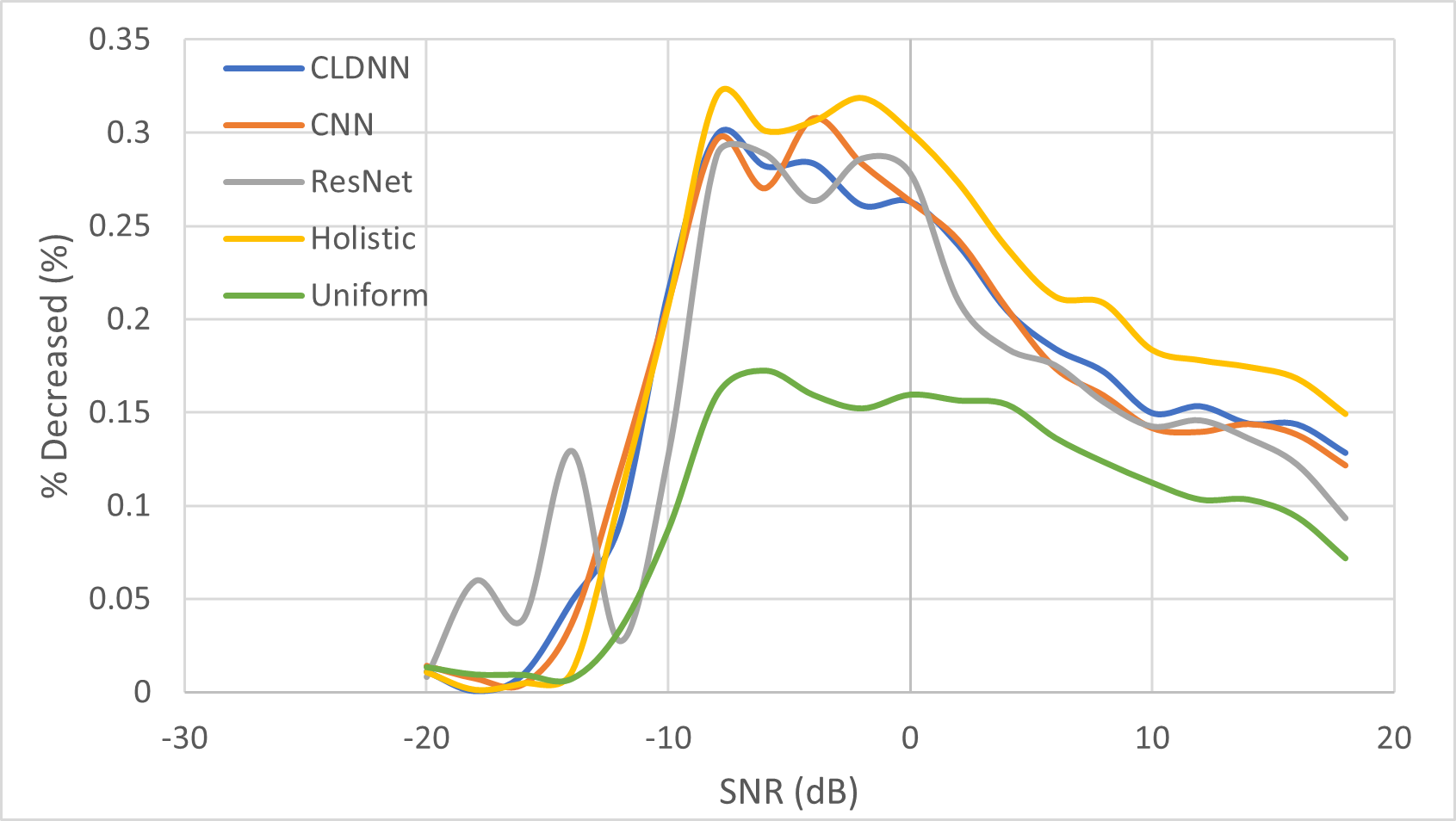}
     \caption{Percent decrease in accuracy due to the black-box adversarial attack vs SNR for 64 sample classifiers.}
     \label{fig:black64pec}
 \end{subfigure}
 \hfill
    \caption{Classification accuracy of different classifiers of 64 input samples with present of \textit{black-box} attack}
    \label{fig:blackbox}
\end{figure}

\begin{table}
\centering\footnotesize
\caption{The average success rate of attack from using each classifier model under the black-box scenario}
\label{table:avgAttack}\medskip
\begin{threeparttable}
\begin{tabular}{P{2.10cm}|P{1.3cm}P{1.3cm}P{1.3cm}} 
\hline
SubsamplerNet & \multicolumn{3}{c}{ Effectiveness of attack } \\
samples & 64 & 32 & 16 \\ \hline
CLDNN & 17.4\% & 15.1\% & 9.9\% \\
CNN & 19.2\% & 15.4\% & 14.5\% \\
ResNet & 17.8\% & 14.3\% & 9.2\% \\
Holistic & 19.4\% & 18.5\% & 8.1\% \\
Uniform & 17.0\% & 11.6\% & 9.1\% \\ \hline
\end{tabular}
  \begin{tablenotes}[para,flushleft]
  This table demonstrates how effective each subsampling scheme is, on average, on reducing the accuracy of \textit{victim}'s model that used different subsampling schemes (black-box senario). The value represent average \% decrease in accuracy caused by each scheme.
  \end{tablenotes}
\end{threeparttable}
\end{table}

\begin{figure}
\centerline{\includegraphics[width=8cm]{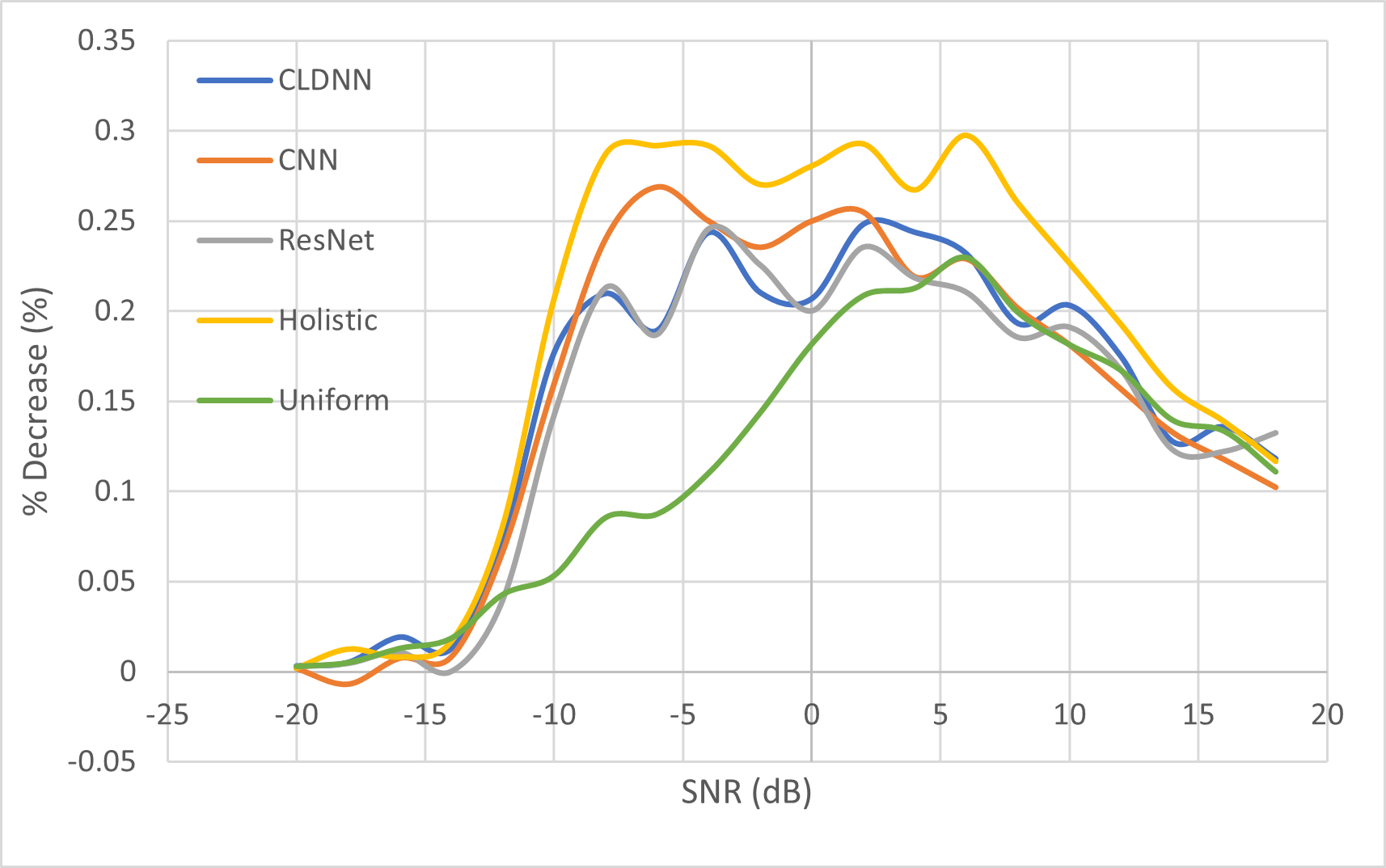}}
\caption{The average \% decrease in accuracy caused by the adversarial example crafted with each subsampling scheme on other 32 sample classifiers vs SNR.}
\label{fig:attacker}
\end{figure}
\begin{table*}[ht]
\centering\footnotesize
\caption{Comparison of accuracy of models trained with whole and selected SNR data, under a white-box attack setting}
\label{table:snrSelect}\medskip
\begin{threeparttable}
\begin{tabular}{P{2cm}P{2cm}P{2cm}|P{2cm}P{2cm}|P{2cm}P{2cm}}
\hline
\multicolumn{1}{c|}{} & \multicolumn{2}{c|}{Accuracy when attacked (64 samples)} & \multicolumn{2}{c|}{Accuracy when attacked (32 samples)} & \multicolumn{2}{c}{Accuracy when attacked (16 samples)} \\
\multicolumn{1}{c|}{Subsampler Net} & whole SNR & selected SNR & whole SNR & selected SNR & whole SNR & selected SNR \\ \hline
\multicolumn{1}{c|}{CLDNN} & 48.0\% & 43.8\% & 41.8\% & 36.8\% & 39.5\% & 33.3\% \\
\multicolumn{1}{c|}{CNN} & 42.9\% & 44.7\% & 43.9\% & 36.6\% & 36.4\% & 32.2\% \\
\multicolumn{1}{c|}{ResNet} & 48.4\% & 52.2\% & 43.7\% & 43.7\% & 40.7\% & 31.4\% \\
\multicolumn{1}{c|}{Holistic} & 50.2\% & 47.5\% & 42.2\% & 40.0\% & 38.2\% & 32.4\% \\
\multicolumn{1}{c|}{Uniform} & 51.0\% & 48.8\% & 51.9\% & 50.9\% & 43.7\% & 35.2\% \\ \hline

\end{tabular}
  \begin{tablenotes}[para,flushleft]
  Classification accuracy of a model that is trained with data consisting of the whole SNR set and a model that is trained with data consisting of only 0dB and 18dB SNR value when it is attacked in a white-box manner.
  \end{tablenotes}
\end{threeparttable}
\end{table*}

\begin{table}
\centering\footnotesize
\caption{Comparison of training time of models trained with whole and selected SNR data}
\label{table:runtime}\medskip
\begin{threeparttable}
\begin{tabular}{ccc}
\hline
\multicolumn{1}{c|}{} & \multicolumn{2}{c}{Average training time (sec)} \\
\multicolumn{1}{c|}{samples} & whole SNR & selected SNR \\ \hline
\multicolumn{1}{c|}{64} & 864 & 170 \\
\multicolumn{1}{c|}{32} & 590 & 115 \\
\multicolumn{1}{c|}{16} & 519 & 106 \\
\hline
\end{tabular}
\end{threeparttable}
\end{table}

\subsection{Best Defense and Attack Strategies}
From Fig. \ref{fig:whitebox} and Fig. \ref{fig:blackbox}, it is obvious that the uniform subsampling scheme provides the best, or very close to the best accuracy and robustness both with and without adversarial attacks, across the studied SNR range. Results in Table \ref{table:avgAcc} and Table \ref{table:percentDec} also show that the uniform subsampling scheme outperforms the rest, on average, at all considered subsampling rates. This suggests that \textbf{the uniform subsampler would be the best choice for the \textit{victim classifier}} in presence of uncertainty about the subsampler assumed by the attacker.
However, it is important to note that the defender can benefit from occasionally using other subsampling strategies to avoid leaking knowledge about the employed strategy to the attacker, since the white-box attack, even when using uniform subsampling by the defender, is very effective compared to black-box scenarios as shown in Table \ref{table:avgAcc}.

To elaborate on the effect of creating uncertainty about the defender's strategy at the attacker, as well as identify the best attack strategy in presence of such uncertainty, we have tested how varying the assumption about the employed subsampler can affect the attack's effectiveness. Fig. \ref{fig:attacker} and Table \ref{table:avgAttack} demonstrate how effective each subsampling scheme used by the attacker was, on average, for reducing the accuracy of the victim's model in a black-box setting. From this result, we can conclude that using uniform subsampling is actually the least effective choice for the attacker in the black-box scenario; to maximize the attacking performance, \textbf{the best the attacker can do is using a Holistic subsampler and a CNN Subsampler Net at high and low subsampling rates, respectively}. 

\subsection{Free Computational Gains with SNR Selective Training}
We further investigate the adversarial setting in a case where we train the classifiers using only data sets of a selected pair of SNR values, instead of the whole 20 provided by RML2016.10b. SNR selection can accelerate training, and can facilitate feasibility when either computational resources are limited or access to multiple SNR values is difficult during data collection. From the results in the previous work \cite{ramjee2019fast}, we can hypothesize that the combination of 0dB and 18dB would give us the best performance, compared to selecting other SNR value pairs, and hence we fix that choice here.  

\begin{figure}[ht]
 \centering
 \begin{subfigure}[a]{8cm}
     \centering
     \includegraphics[width=8cm]{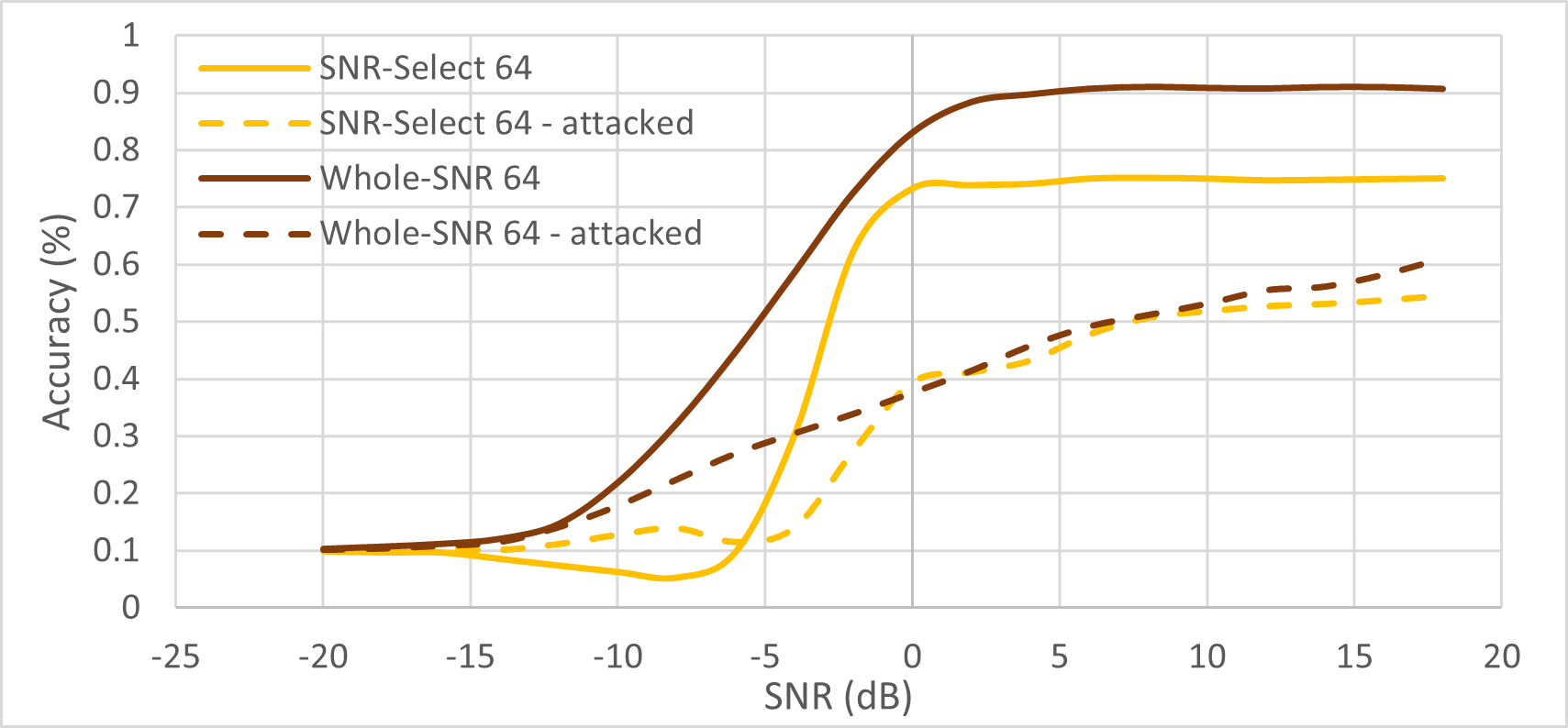}
     \caption{64 samples}
     \label{fig:y equals x}
 \end{subfigure}
 \hfill
 \begin{subfigure}[b]{8cm}
     \centering
     \includegraphics[width=8cm]{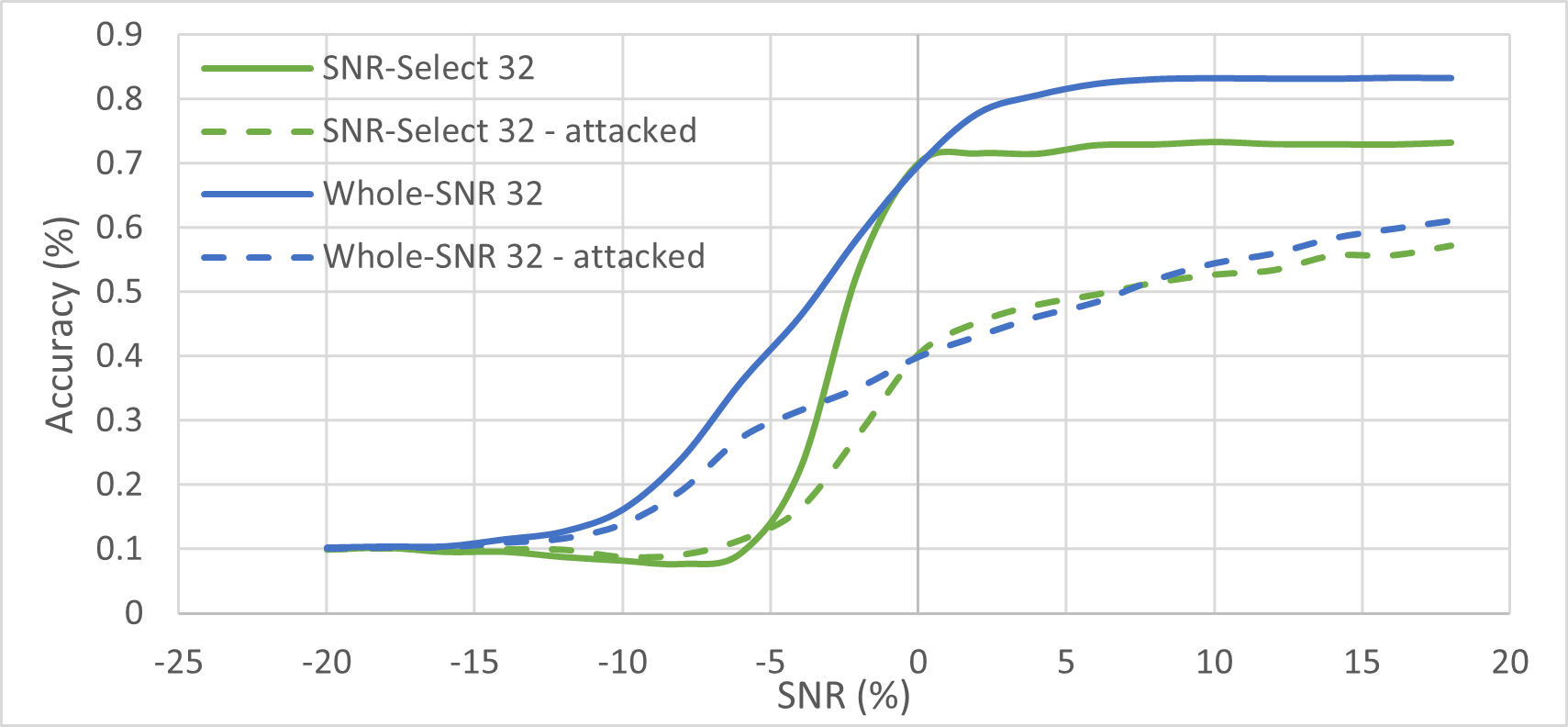}
     \caption{32 samples}
     \label{fig:three sin x}
 \end{subfigure}
 \hfill
 \begin{subfigure}[c]{8cm}
     \centering
     \includegraphics[width=8cm]{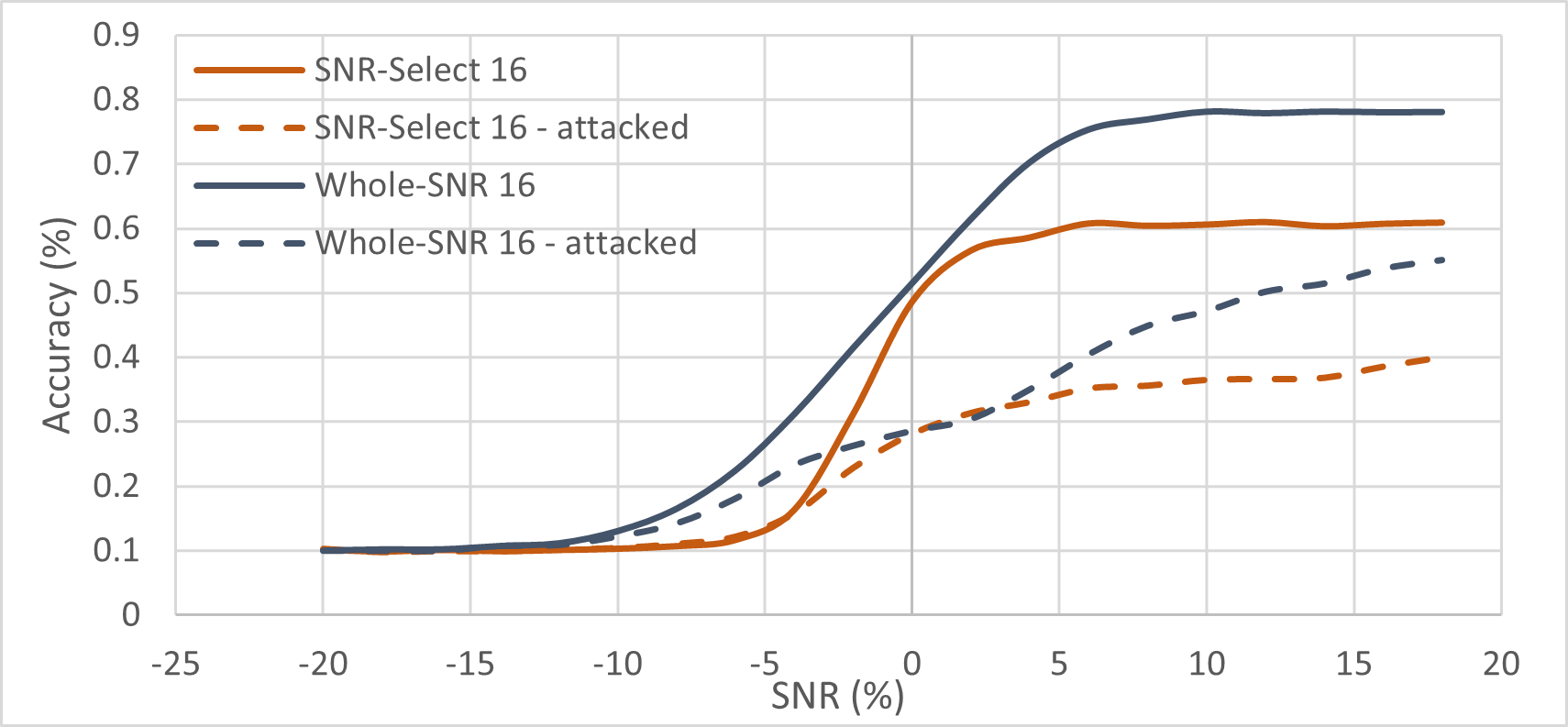}
     \caption{16 samples}
     \label{fig:five over x}
 \end{subfigure}
    \caption{Accuracy vs SNR for classifiers trained with whole/selected SNR data before and after white-box attack.}
    \label{fig:snrSelect}
\end{figure}

As shown in Fig. \ref{fig:snrSelect}, \textbf{the classifier with SNR selective training shows similar accuracy as the traditional classifier in an adversarial setting} at SNR values over 0 dB. Further, Table \ref{table:snrSelect} shows that in some cases, like using CNN or ResNeT subsamplers with a subsampling rate of $\frac{1}{2}$, SNR selective training led to performance improvements in an adversarial setting. However, this observation only holds down to a certain subsampling rate. Evident from the result for the classifier with 16-sample inputs, if the subsampling rate is too aggressive, SNR selective training leads to performance degradation in presence of an attack. 

To emphasize the low computational cost of SNR selective training, we show the training times in Table \ref{table:runtime}. In summary, even though SNR selective training has been shown previously to reduce training time at the cost of performance degradation, \textbf{we identified scenarios in an adversarial setting with mild subsampling rates, where this computational gain comes at no cost in performance}. 

\section{Conclusion and Future Work}
We studied the efficacy of gradient-based adversarial attacks on deep learning models that utilize data-driven subsampling for wireless modulation classification. We highlighted how knowledge of the subsampling scheme employed at both the victim model, as well as at the attacker to generate the perturbation, is crucial for the success of the defender and attacker, respectively. With lack of such knowledge, we identified the best defense and attack strategies. We further identified opportunities for saving computational resources in an adversarial setting at no cost in performance via SNR selective training. We believe that this work lays the ground for further investigations on data-driven subsampling strategies for wireless communication systems employing deep learning in an adversarial setting. We are particularly interested in studying the impact of SNR knowledge in such settings. 

\bibliography{ref}
\bibliographystyle{IEEEtran}

\end{document}